\documentclass[12pt]{iopart}

\usepackage{iopams} 
\usepackage{graphicx}
\usepackage{caption}
\usepackage{revsymb}
\usepackage{bm}
\usepackage{xcolor}
\usepackage{amsmath}

\newcommand{\HIJ}{Helmholtz Institute Jena, Fröbelstieg 3, 07743 Jena, Germany} 
\newcommand{\GSI}{GSI Helmholtzzentrum für Schwerionenforschung GmbH, Planckstraße 1, 64291 Darmstadt, Germany}
\newcommand{\IOQ}{Institute of Optics and Quantum Electronics, Max-Wien-Platz 1, 07743 Jena, Germany}

\begin{document}

\title{Bright polarised x-ray flashes from dense plasmas}

\author{Q. Qian}
\address{G\'{e}rard Mourou Center for Ultrafast Optical Science, University of Michigan, 2200 Bonisteel Boulevard, Ann Arbor, Michigan 48109, USA}

\author{C.P. Ridgers}
\address{York Plasma Institute, Department of Physics, University of York, York, YO10 5DD, United Kingdom}

\author{S.V. Bulanov}
\address{ELI Beamlines Facility, Extreme Light Infrastructure ERIC, Za Radnicí 835, 25241 Dolní Břežany, Czech Republic}

\author{T. Grismayer}
\address{GoLP/Instituto de Plasmas e Fusão Nuclear, Instituto Superior Técnico, Universidade de Lisboa, 1049-001 Lisbon, Portugal}

\author{P. Hadjisolomou}
\address{ELI Beamlines Facility, Extreme Light Infrastructure ERIC, Za Radnicí 835, 25241 Dolní Břežany, Czech Republic}

\author{D. Seipt}
 \address{\HIJ}
 \address{\GSI}
 \address{\IOQ}

\author{M. Vranic}
\address{GoLP/Instituto de Plasmas e Fusão Nuclear, Instituto Superior Técnico, Universidade de Lisboa, 1049-001 Lisbon, Portugal}

\author{A.G.R. Thomas}
\address{G\'{e}rard Mourou Center for Ultrafast Optical Science, University of Michigan, 2200 Bonisteel Boulevard, Ann Arbor, Michigan 48109, USA}

\ead{qqbruce@umich.edu}
\vspace{10pt}

\begin{abstract}
Creating a plasma dominated by strong-field QED (SFQED) effects is a major goal of new multi-PW laser facilities.  This is motivated by the fact that the fundamental dynamics of such plasmas is poorly understood and plays an important role in the electrodynamics of extreme astrophysical environments such as pulsar magnetospheres.  The most obvious observable for which such a regime has been reached is the production of a bright flash of x-rays, but distinguishing this from other sources of hard x-rays (e.g., bremsstrahlung) is a major challenge.  Here we show that the photons from the X-ray flash are highly polarised, as compared to the unpolarised background, i.e., polarisation is an indicator that the SFQED plasma has really produced.  For a laser of intensity $10^{21}$\thinspace Wcm$^{-2}$ impinging on a solid Al target, the photons of the flash with energy $>10$\thinspace keV are $>65\%$ polarised.  
%At higher intensity, $5\times10^{21}$\thinspace Wcm$^{-2}$, $>1$\thinspace MeV photons are almost completely polarised. 
\end{abstract}

\section{Introduction}

New multi-PW power lasers are on the verge of creating an entirely new state in the laboratory dominated by the interplay of strong-field quantum electrodynamics (QED) processes and ultra-relativistic plasma effects \cite{Ridgers_12, Ridgers_12_2, Fedotov_PR_2023, GonoskovA_RevModphy_2022, HadjisolomouP_POP_2023}. These ‘QED-plasmas’ are found in extreme astrophysical environments, such as pulsar magnetospheres \cite{Timokhin_10, Cruz_AJL_2021}, but have yet to be realised in the laboratory. In this paper, we will describe a possible way to observe this behaviour by observing the bright flashes of hard photons, which are a hallmark of QED plasmas, using polarimetry.  The two dominant processes in laser-generated QED-plasmas are hard photon emission as electrons (and positrons) in the plasma are accelerated by the strong laser and plasma electromagnetic fields (nonlinear inverse Compton scattering - NLCS).  This process has recently been experimentally demonstrated using all-optic setups 
\cite{Ta_NP_2012, Sarri_PRL_2014, Poder_PRX_2018, Cole_PRX_2018, Pirozhkov_2024}, and the generation of gamma-rays with energies up to GeV has been reported \cite{Mirzaie_NP_2024}. The hard photons can then decay to electron-positron pairs in these fields (by the multiphoton Breit-Wheeler process) \cite{Bell_08,Fedotov_10}.  Very many of these interactions occur in the plasma.  Simplified models describing these strong-field QED processes in the arbitrary electromagnetic fields in the plasma, and capable of dealing with the high multiplicity of the interactions, have been developed \cite{Kirk_10} and included in particle-in-cell (PIC) plasma modelling codes \cite{Elkina_11,Ridgers_14,Gonoskov_15, Grismayer_POP_2016}.   PIC simulations have been used to explore the interplay of strong-field QED and ultra-relativistic plasma processes in QED-plasmas in both the laboratory (examples in the following Refs. \cite{Nerush_10,Brady_13,Capdessus_18,Liu_17,Grismayer_17,Slade_Lowther_19,Liseykina_21}) and astrophysical settings \cite{Timokhin_10,Timokhin_15,Chernoglazov_24}.  In both cases, substantial energy is emitted as hard photons, which can then drive an electron-positron cascade \cite{Seipt_NJP_2021}.  In the laboratory case the radiation reaction, resulting from the emission of hard photons, can cause almost all of the laser energy to be absorbed \cite{Zhang_15}, strongly modifying the plasma dynamics (for example, quenching radiation pressure ion acceleration \cite{Kirk_13, Del_Sorbo_18}).  Most recently, these models have been extended to include the spin of electrons and positrons in the plasma and the polarisation of the hard photons \cite{Del_Sorbo_17, Del_Sorbo_17_2, Seipt_18, Li_19, Seipt_PRA_2020, Torgrimsson_NJP_2021, Seipt_NJP_2021, Seipt_PPCF_2025}.  Spin and polarisation have been included in some PIC codes \cite{Wan_23, Qian_23, Zhu_24}, following the well established methodology used in the particle physics code {\small{CAIN}} \cite{Chen_1995_CAIN}.

The burst of hard photons is therefore one of the observables of strong-field QED effects in a plasma. The brilliant NLCS ‘x-ray flash’\footnote{Note that the hard photons are often referred to as gamma-rays for photon energies $>1$\thinspace MeV. We will refer to all hard photons as X-rays for simplicity. This is because we consider hard photons with 1-100\thinspace keV energies as well as those above 1\thinspace MeV.}\cite{Ridgers_12,Nakamura_12} is predicted to be one of the most prominent signatures that the QED-plasma regime has been reached and thus a first unambiguous observation of this is a high priority \cite{ELI_NP,Altarelli_19,Doria_20,DiPiazza_22}. Several groups have reported an observation of the NLCS x-ray flash but the background was very high \cite{Shou_23,Pirozhkov_24}. The main source of background x-ray emission in the $>$keV energy range is from bremsstrahlung emission as electrons in the laser-produced plasma scatter from the ions. The bremsstrahlung occurs over a much longer timescale than the NLCS (tens of ps compared to 100\thinspace fs), and so it is difficult to model using PIC. Recent simulations using a hybrid-PIC model have simulated the bremsstrahlung for the required 100\thinspace ps \cite{Morris_21} and shown that the background is much higher than expected from previous, shorter timescale PIC simulations \cite{Vyskočil_18}. Hybrid-PIC has recently been used to show that NLCS may be observed clearly above bremsstrahlung only at intensities $>10^{22}$\thinspace Wcm$^{-2}$ \cite{Ingleby_25}, and within a certain range of target thickness \cite{Martinez_PRR_2020}.  Observing the x-ray flash at lower intensities more commonly available in experiments ($\sim 10^{21}$\thinspace Wcm$^{-2}$) is therefore a challenge. To overcome this, we focus on identifying unique signatures of the NLCS process. In particular, polarisation is a promising observable. Recently, the measurement of x-ray polarisation has become increasingly important in fields ranging from atomic physics \cite{Go_SR_2024} and astrophysics \cite{Ilie_Astro_Pac_2019} to laser-plasma interactions \cite{Schnell_NC_2013}, with measurement capabilities spanning a few keV up to GeV-scale photon energies.

In this paper, we will show that the NLCS flash gives rise to highly polarised x-rays (50-70\%) in the energy range 1-10\thinspace keV.  And for the moderate laser intensity region we studied here ($I\sim 10^{21}$\thinspace Wcm$^{-2}$), the polarisation signal increases with the photon energy and can reach above $80\%$. This demonstrates that x-ray polarimetry is an ideal tool for observing the x-ray flash and the onset of the QED-plasma regime. % The photon generated through bremsstrahlung, which in general is unpolarised or only weakly polarised, will serve as the background for this measurement. The bremsstrahlung signal can actually be larger than the NLCS signal, but we can control it by varying the target thickness and atomic number. The directionality of the bremsstrahlung photon can be modified by changing the target shape, and this can be useful for reducing the bremsstrahlung background \cite{Morris_21}.  
 It is worth mentioning that the bremsstrahlung process can also generate polarised X-rays when measured at some specific angles \cite{Shohet_PRL_1977}. The polarised light from bremsstrahlung is anisotropic, while the polarised light from NLCS is basically isotropic. One can easily rule out polarised X-rays from the bremsstrahlung effect by taking several measurements around the targets. The paper will be structured as follows.  First, the polarised NLCS emission rates will be introduced, and a simplified analytical model for the expected degree of polarisation of the x-ray flash will be developed.  Next the implementation of the polarised emission algorithm in the PIC code {\small{OSIRIS}} will be discussed.  {\small{OSIRIS}} simulation results will then be compared to the simple model and used to show that the x-ray photons are indeed highly polarised.

\section{Model for hard photon polarisation}

\subsection{Polarised photon emission in laser-plasma interactions}

Hard photon emission can be described using the Locally-constant Crossed Field Approximation (LCFA) \cite{Ridgers_17, Bell_08}.  We will briefly summarise this model here for convenience.  We make two key assumptions. 

\begin{enumerate}
    \item {The photon formation length is small, valid for $a_0\gg 1$.}
    \item{The electric field $E_{L}$ in the plasma is much less than the critical (Schwinger) field $E_S$.}
\end{enumerate}

Here, $a_0=eE_L/m_ec\omega_L$ is the strength parameter and depends on the electric field of the laser $E_L$ and its frequency $\omega_L$. $E_S$ is given by $E_S=m_e^2c^3/e\hbar$, is defined in terms of the electron rest mass $m_e$, the speed of light $c$, the elementary charge $e$, and the reduced Planck constant $\hbar$. The first assumption means that the emission can be considered to occur locally and thus only depends on the local electromagnetic fields.  The second assumption means that the emission only depends on the invariant $\chi_e=E_{RF}/E_S$ -- the quantum efficiency parameter.  As a result, the photon emission rate in the electromagnetic fields in the plasma is the same as in any convenient configuration of electromagnetic fields with the same value of $\chi_e$.  We choose a plane wave, i.e., crossed electric and magnetic fields, under the previous assumption of local emission.  As a result, the emission in the arbitrary electric and magnetic fields in the plasma can be well described by NLCS. Notice that the QED rates can be modified for an interaction with a tightly focused laser pulse \cite{Ilderton_PRA_2019, Piazza_PRL_2016}. However, for the configuration explored in this study, the laser spot focal spot size is moderate ($w_0 \sim 4\lambda$), so we don't need to consider this correction. 

Describing the polarisation of the emitted photons requires choosing a basis for this polarisation.  In the LCFA model, we assume that the electric and magnetic fields are perpendicular to each other in the electron's rest frame.  Thus, we can define a mutually orthogonal basis $(\hat{\pmb{\alpha}},\,\hat{\pmb{\beta}},\,\hat{\pmb{\kappa}})$, where $\hat{\pmb\alpha}=\pmb{E}_{RF}/|\pmb{E}_{RF}|$, $\hat{\pmb\beta}$ is the equivalent for the magnetic field in the rest frame $\hat{\pmb\beta}=\pmb{B}_{RF}/|\pmb{B}_{RF}|$, and $\hat{\pmb{\kappa}}$ points along the electron momentum.
 
The photon polarisation state can be represented as $\hat{\pmb \psi} = a_1 \hat{\pmb{\psi}}_1 + a_2 \hat{\pmb{\psi}}_2$ . which the vector $\hat{\pmb{\psi}}_1 = [\pmb{{E}}-\hat{\pmb{k}}\cdot (\hat{\pmb{k}} \cdot \pmb{{E}})+ (\hat{\pmb{k}} \times \pmb{{B}})]/|\pmb{{E}}-\hat{\pmb{k}}\cdot (\hat{\pmb{k}} \cdot \pmb{{E}})+ (\hat{\pmb{k}} \times \pmb{{B}})|$, and $\hat{\pmb{\psi}}_2 = [\pmb{{B}}-\hat{\pmb{n}}\cdot (\hat{\pmb{n}} \cdot \pmb{{B}})- (\hat{\pmb{n}} \times \pmb{{E}})]/|\pmb{{B}}-\hat{\pmb{n}}\cdot (\hat{\pmb{k}} \cdot \pmb{{B}})+ (\hat{\pmb{k}} \times \pmb{{E}})|$. Here, $\hat{\pmb{k}}$ is the unit vector along the photon propagation direction, and $\pmb{E}$ and $\pmb{B}$ are the local electric and magnetic fields experienced by the photon.At the moment of emission, $\hat{\pmb{\psi}}_1 = \hat{\pmb{\alpha}}$ and $\hat{\pmb{\psi}}_2 = \hat{\pmb{\beta}}$, coinciding respectively with the directions of the rest frame electric and magnetic fields experienced by the parent lepton. Under the collinear emission approximation ($\gamma \gg 1$), the photon propagation direction $\hat{\pmb{k}}$ is aligned with the initial lepton momentum, i.e., $\hat{\pmb{\kappa}}$. The stokes vector $\pmb{\xi} = (\xi_1,\xi_2,\xi_3)$ correspond to this basis is  $\xi_1 = a_1a_2^* + a_1^*a_2$, $\xi_2 = i(a_1a_2^*-a_1^*a_2)$, $\xi_3 = |a_1|^2-|a_2|^2$. For observation, however, the relevant basis may differ due to the various field configurations the photon encounters after emission. In general, the observational basis can be expressed as: 
\begin{equation} 
\hat{\pmb  o}_1 = \hat{\pmb \psi}_1\cos(\phi) + \hat{\pmb \psi}_2\sin(\phi),\  
\hat{\pmb  o}_2 = - \hat{\pmb \psi}_1\sin(\phi) + \hat{\pmb \psi}_2\cos(\phi)
\end{equation}

The Stokes parameter under the new observation basis $\pmb\xi^{(o)} $ is then:
\begin{equation} 
\xi_1^{(o)}= \xi_1\cos(2\phi) - \xi_3\sin(2\phi),\ 
\xi_2^{(o)} = \xi_2,\  
\xi_3^{(o)} = \xi_1\sin(2\phi) +\xi_3\cos(2\phi)
\label{eq:observe_stokes_parameter}
\end{equation}

% For example, the calculation of polarisation resolved pair production process can be treated as a mesurement, which the basis of of this measure is $(\hat{\pmb{\psi}}_1’, \hat{\pmb{\psi}}_2’, \hat{\pmb{k}})$. The vector $\hat{\pmb{\psi}}_1’ = [\pmb{{E}}-\hat{\pmb{k}}\cdot (\hat{\pmb{k}} \cdot \pmb{{E}})+ (\hat{\pmb{k}} \times \pmb{{B}})]/|\pmb{{E}}-\hat{\pmb{k}}\cdot (\hat{\pmb{k}} \cdot \pmb{{E}})+ (\hat{\pmb{k}} \times \pmb{{B}})|$, and $\hat{\pmb{\psi}}_2’ = [\pmb{{B}}-\hat{\pmb{n}}\cdot (\hat{\pmb{n}} \cdot \pmb{{B}})- (\hat{\pmb{n}} \times \pmb{{E}})]/|\pmb{{B}}-\hat{\pmb{n}}\cdot (\hat{\pmb{n}} \cdot \pmb{{B}})+ (\hat{\pmb{n}} \times \pmb{{E}})|$. Here, $\hat{\pmb k}$ represents the direction of propagation of the photon,  and $\pmb{E}$ and $\pmb{B}$ are the electric and magnetic fields experienced by the photon at the pair-production event.

In the actual code implementation, the Stokes parameters at the moment of photon emission can be obtained directly from the polarization-resolved NLCS rate calculation. However, to analyze the photon’s polarization at the point of observation—when the photon interacts with external electromagnetic fields—it is necessary to transform the Stokes parameters from the emission frame to the observation frame. This transformation requires not only the value of the Stokes parameters at emission but also information about the polarization basis at that instant. To minimize computational memory usage, we seek to reduce the number of parameters required to fully specify the photon polarization state. We find that it is sufficient to represent the polarization state by a single vector quantity, $\pmb{P} = (P_{\psi_1}$, $P_{\psi_2}$, $P_{{k}})$. Which $P_{\psi_1}$, $P_{\psi_2}$, $P_{{k}}$ is the component in the $({\hat{\pmb\psi}_1}, {\hat{\pmb\psi}_2}, {\hat{\pmb k}})$ coordinate. The relationship between $P_{\varepsilon}$, $P_{k}$, $P_{\beta}$ and the Stokes parameter can be written as:
\begin{equation}
	\xi_1 = \frac{2P_{\psi_1} P_{\psi_2}}{\sqrt{1-P_{{k}}^2}},\ \ 
	\xi_2 = P_{{k}},\ \ \xi_3 = \frac{P_{\psi_1}^2 -  P_{\psi_2}^2}{\sqrt{1-P_{{k}}^2}}
\label{eq:pvector_stokes_relation}
\end{equation}

Under the observation frame, we can obtain the observed Stokes parameters by project vector $\pmb P$ to the observation basis $(\hat{\pmb o}_1, \hat{\pmb o}_2, \hat{\pmb k})$.
\begin{equation}
 {P}_{{ o_1}} = \pmb{P} \cdot \hat{\pmb o}_1 =  {P}_{\psi_1} \cos(\phi) +   {P}_{\psi_2} \sin(\phi)
\end{equation}
\begin{equation}
P_{{o_2}} = \pmb{P} \cdot \hat{\pmb o}_2= -P_{\psi_1} \sin(\phi) + P_{\psi_2} \cos(\phi)
\end{equation}

Because the photon momentum direction does not change during propagation, $\xi_2^{(o)} = P_{{k}} = \xi_2$ remains unchanged. The Stokes parameter under the observation basis, calculated using Eq.~\ref{eq:pvector_stokes_relation} can be written as:

\begin{equation}
	\xi_{1}^{(o)} = -\sin(2\phi)\xi_3 + \cos(2\phi)\xi_1
, \ \ \xi_{3}^{(o)} = \cos(2\phi)\xi_3 + \sin(2\phi)\xi_1
\end{equation}

This relation is equivalent to Eq.~\ref{eq:observe_stokes_parameter}. Consequently, the full photon polarisation state can be represented by the reconstructed vector $\pmb{P}$, with the appropriate transformation applied as outlined in Eq.~\ref{eq:pvector_stokes_relation}. Throughout the remainder of this work, the photon polarisation is measured in the basis $(\hat{\pmb{o}}_1,\, \hat{\pmb{o}}_2,\, \hat{\pmb{k}})$, where $\hat{\pmb{o}}_1$ denotes the polarization direction of the linearly polarised laser, $\hat{\pmb{k}}$ is the photon propagation (momentum) direction, and $\hat{\pmb{o}}_2$ is defined as $\hat{\pmb{o}}_2 = \hat{\pmb{o}}_1 \times \hat{\pmb{k}}$.

With this polarisation basis properly defined, we present below the spin- and polarization-resolved radiation spectrum in the LCFA approximation~\cite{Torgrimsson_NJP_2021}:

\begin{equation}
F_{\text{NLC}}(\chi_e, \textbf{s}_i,\textbf{s}_f,  \varepsilon_\gamma, \bm{\xi}) = F_0 + \bm{\xi} \cdot \textbf{F} = F_0 + \xi_1F_1 + \xi_2F_2 + \xi_3F_3,
\label{eq:F_NLCS}
\end{equation}

where $\textbf{s}_i$, $\textbf{s}_f$ are the initial and final spin state of the lepton, $\varepsilon_\gamma$ is the radiated photon energy, $\bm{\xi}$ is the Stokes vector for the photon. The rate of the NLCS radiation process can be written as:
% \begin{equation}
% \bm{\xi}_\gamma = \xi_1\pmb{\hat\epsilon}_{EB} + \xi_2\pmb{\hat\epsilon}_{2} + \xi_3\pmb{\hat\epsilon}_{E}.
% \end{equation} 

\begin{equation}
R_{\text{NLC}}= C_0\int_0^1 d\lambda F_{\text{NLC}}
\end{equation}

 where $\lambda = {\varepsilon_\gamma}/{\varepsilon}$ is the momentum transfer ratio between the initial energy of the lepton $\varepsilon$ and the radiated photon $\varepsilon_\gamma$. The prefactor $C_0 = \frac{\alpha}{4b_p}$, where $\alpha = e^2/4\pi$ is the fine structure constant. $b_p = p\cdot \kappa/m^2$ is the quantum energy parameter, which $p$ is the momentum of the lepton before radiation, and $\kappa$ is the wave vector of the colliding laser. We look at each term of $F_{\text{NLC}}$. The first term on the right-hand side of Eq.~\ref{eq:F_NLCS}, the spin-dependent unpolarised spectrum $F_0$ can be written as:

\begin{equation}
\begin{split}
F_0 & = -\text{Ai}_1(z) - u \frac{\text{Ai}'(z)}{z} + \lambda\frac{\text{Ai}(z)}{\sqrt{z}} ({\textbf{s}}_i \cdot \hat{\pmb{\beta}})  \\  & \left. + \frac{\lambda}{1-\lambda}\frac{\text{Ai}(z)}{\sqrt{z}} ({\textbf{s}}_f \cdot \pmb{\hat{\beta}}) 
- \left( \text{Ai}_1(z) + 2\frac{\text{Ai}'(z)}{z}\right)({\textbf{s}}_f \cdot {\textbf{s}}_i) \right.
 \\ &
 - (u-2)\left(\text{Ai}_1(z) +\frac{\text{Ai}'(z)}{z}\right)({\textbf{s}}_i\cdot \hat{\pmb{\kappa}})({\textbf{s}}_f \cdot\hat{\pmb{\kappa}}).
\end{split}
\label{eq:NLC_stokes_F0}
\end{equation}

Here, $\mathrm{Ai}(z) = \frac{1}{\pi}\int_0^{\infty}\cos\left(\frac{t^3}{3}+zt\right)dt$ is the Airy function, and $\mathrm{Ai}'$, $\mathrm{Ai}_1$ are it's derivative and integral. The argument of the Airy function $z=(r/\chi_e)^{2/3}$ depend on the quantum parameter $\chi_e$, and $r = \lambda/(1-\lambda)$. $u = 1/(1-\lambda) + (1-\lambda)$, $\tilde u = 1/(1-\lambda) - (1-\lambda)$. 
Notice that the spin-resolved radiation rate can be written as: 

\begin{equation}
F_0 =w + \textbf{s}_f \cdot \textbf{g}.
\end{equation}

\begin{equation}
    w =-\text{Ai}_1(z) - u \frac{\text{Ai}'(z)}{z} + \lambda\frac{\text{Ai}(z)}{\sqrt{z}} ({\textbf{s}}_i \cdot \pmb{\hat{\beta}}),
\end{equation}

\begin{equation}
\begin{split}
    \textbf{g} = & \frac{\lambda}{1-\lambda}\frac{\text{Ai}(z)}{\sqrt{z}} \pmb{\hat{\beta}}
- \left( \text{Ai}_1(z) + 2\frac{\text{Ai}'(z)}{z}\right) {\textbf{s}}_i
 \\ &
 - (u-2)\left(\text{Ai}_1(z) +\frac{\text{Ai}'(z)}{z}\right)({\textbf{s}}_i\cdot\pmb{\hat{\kappa}}) \pmb{\hat{\kappa}}
\end{split}
\end{equation}

Which $w$ is the unpolarised, spin-averaged NLCS rate. We can also obtain the expected spin polarisation vector:
\begin{equation}
\textbf{S}_{SQA} = \frac{\textbf{g}}{w}.
\label{eq:SQA_spin}
\end{equation}

We then look at the photon polarisation-related radiation rate $\textbf{F} = (F_1, F_2, F_3)$:

\begin{equation}
 \begin{split}
F_1  =  &-\frac{\text{Ai}(z)}{\sqrt{z}}\left[\frac{\lambda}{1-\lambda} (\textbf{s}_i\cdot \pmb{\hat\alpha}) +\lambda(\textbf{s}_f\cdot \pmb{\hat\alpha})\right]
\\ &  - \frac{\lambda^2}{2(1-\lambda)}\text{Ai}_1(z)\left[(\textbf{s}_f \cdot \pmb{\hat\alpha})(\textbf{s}_i \cdot \pmb{\hat\beta})+(\textbf{s}_f \cdot \pmb{\hat\beta})(\textbf{s}_i \cdot \pmb{\hat\alpha})\right] \\&  - \frac{\tilde u}{2}\frac{\text{Ai}'(z)}{z}(\textbf{s}_{f}\times\textbf{s}_{i})\cdot\pmb{\hat\kappa}.
 \end{split}
\label{eq:NLC_stokes_F1}
\end{equation}

\begin{equation}
 \begin{split}
 F_2  = & -\lambda\left[ \text{Ai}_1(z)+\left(1+\frac{1}{1-\lambda}\right)\frac{\text{Ai}'(z)}{z}\right](\textbf{s}_i\cdot \pmb{\hat{\kappa}})  \\ & -\frac{\lambda}{1-\lambda}\left[ \text{Ai}_1(z)+\left(2-\lambda\right)\frac{\text{Ai}'(z)}{z}\right](\textbf{s}_f\cdot \pmb{\hat{\kappa}}) 
 \\ & 
 +\frac{\lambda^2}{2(1-\lambda)}
 \frac{\text{Ai}(z)}{\sqrt{z}}\left[(\textbf{s}_f\cdot \pmb{\hat{\kappa}})(\textbf{s}_i\cdot \pmb{\hat\beta}) + (\textbf{s}_f\cdot \pmb{\hat\beta})(\textbf{s}_i\cdot \pmb{\hat{\kappa}})\right]
\end{split}
\label{eq:NLC_stokes_F2}
\end{equation}

\begin{equation}
 \begin{split}
 F_3  &= -\frac{\text{Ai}'(z)}{z}+\frac{\text{Ai}(z)}{\sqrt{z}}\left[\frac{\lambda}{1-\lambda} (\textbf{s}_i\cdot \pmb{\hat\beta}) +\lambda(\textbf{s}_f\cdot \pmb{\hat\beta})\right]
 \\ &  - \frac{\lambda^2}{2(1-\lambda)}\text{Ai}_1(z)\left[(\textbf{s}_f \cdot \pmb{\hat\alpha})(\textbf{s}_i \cdot \pmb{\hat\alpha})-(\textbf{s}_f \cdot \pmb{\hat\beta})(\textbf{s}_i \cdot \pmb{\hat\beta})\right]\\
 & -  \frac{u}{2}\frac{\text{Ai}'(z)}{z}(\textbf{s}_f\cdot \textbf{s}_i) -  \left(1-\frac{u}{2}
 \right)\frac{\text{Ai}'(z)}{z}(\textbf{s}_f\cdot \pmb{\hat\kappa})(\textbf{s}_i\cdot \pmb{\hat\kappa}).
\end{split}
\label{eq:NLC_stokes_F3}
\end{equation}

Based on the spectrum, we can calculate the expected photon Stokes vector for a certain emission event: $\langle \pmb \xi\rangle = (\langle\xi_1\rangle, \langle\xi_2\rangle, \langle\xi_3\rangle) = \textbf{F}/F_0$, which $\langle\xi_1\rangle = F_1/F_0$, $\langle\xi_2\rangle = F_2/F_0$, $\langle\xi_3\rangle = F_3/F_0$.  

In the simplest scenario, we consider an unpolarised particle colliding with an infinitely extended, linearly polarised plane wave, neglecting any energy loss due to radiation reaction. The quantum parameter experienced by the particle is denoted by $\chi_e$. In this regime, the emitted photon can only be linearly polarised. The photon polarisation spectrum, $\xi_3(\omega)$, is then given by
\begin{equation}
\xi_3(\omega) = \frac{F_3}{F_0} = \frac{\mathrm{Ai}'(z)}{z\,\mathrm{Ai}_1(z) + u\,\mathrm{Ai}'(z)},
\end{equation}

By applying the asymptotic expansions of the Airy function and its derivative, we obtain the limiting behavior of $\xi_3(\omega)$ for small and large $z$:
\begin{equation}
\xi_3(\omega) \approx \frac{\Gamma\left(\frac{2}{3}\right) \, 3^{2/3}}
{\frac{\pi\sqrt{3}}{3}z + u\,\Gamma\left(\frac{2}{3}\right) 3^{2/3}}
\approx \frac{1}{u} + \frac{1.28z}{u^2}, \qquad \text{as } z \rightarrow 0,
\label{eq:low_z_limit}
\end{equation}
\begin{equation}
\xi_3(\omega) \approx \frac{1}{u - 1}, \qquad \text{as } z \rightarrow +\infty,
\label{eq:high_z_limit}
\end{equation}
where $z = \left[\frac{\lambda}{\chi_e (1-\lambda)}\right]^{2/3}$. Note that the polarisation spectrum approaches Eq.~\eqref{eq:low_z_limit} when $\chi_e \gg 1$, or equivalently as $\lambda \rightarrow 0$. In this low-energy photon limit, Eq.~\eqref{eq:low_z_limit} predicts a degree of polarisation of approximately $50\%$. In contrast, when $\chi_e \ll 1$, the polarisation spectrum is better described by Eq.~\eqref{eq:high_z_limit}, or when $\lambda \rightarrow 1$.

In summary, the agreement with Eq.~\eqref{eq:low_z_limit} improves for large values of $\chi_e$, while Eq.~\eqref{eq:high_z_limit} becomes more accurate as $\chi_e$ decreases.

\subsection{Model prediction for degree of polarisation}\label{simple_model}

We seek to predict the expected degree of polarization of hard photons emitted during laser-plasma interactions. Here, we develop a simple model to achieve this, focusing on the widely used configuration of a laser interacting with a slab target. Specifically, we consider a fully ionized aluminum target with electron density $n_e = 450\, n_c$, where $n_c = 1.75 \times 10^{21}\,\mathrm{cm}^{-3}$ is the critical density for $800\,\mathrm{nm}$ laser light. The plasma skin depth for this solid-density target is given by $\delta_{p} = \frac{\omega_{pe}}{c} = \frac{n_c}{n_e}\lambda$, where $\lambda$ is the laser wavelength. At relativistic laser intensities, it is more appropriate to use the relativistic skin depth, which can be approximated as: $\delta_{p}^{r} \approx \frac{a_0n_c}{n_e}\lambda$. where $a_0$ is the normalized vector potential of the laser. The number of electrons interacting with the laser is then estimated as $N_e = n_e \pi w_0^2 \delta_{p}^r$, where $w_0$ is the laser spot size. We assume a linearly polarized laser with wavelength $\lambda = 800\,\mathrm{nm}$ and temporal duration $\tau_0 = 20\,\mathrm{fs}$. For simplicity, we further assume that the intensity is uniform within the laser spot size $w_0$. In our model, $w_0$ is chosen to be $\sqrt{2}$ smaller than the spot size used in the particle-in-cell (PIC) simulations, which typically assume a transverse Gaussian profile, so that the total laser power at focus remains consistent. The quantum parameter $\chi_e$ reached by electrons in the plasma can be estimated using~\cite{Ridgers_17}:

\begin{equation}
\chi_e = 0.1 \sqrt{\frac{I}{10^{21}\mbox{\thinspace Wcm}^{-2}}}\frac{\varepsilon}{500 \mbox{\thinspace MeV}}. 
\end{equation}

\noindent where $I$ is the laser intensity and $\varepsilon$ is the electron energy. The intensity $I$ is related to $a_0$ by $I \sim a_0^2 \times 10^{18}\,\mathrm{W\,cm}^{-2} / [\lambda\,(\mu\mathrm{m})]^2$. Assuming that $\varepsilon = a_0 m_e c^2$, the quantum parameter $\chi_e$ depends only on $a_0$.

We then solve for the evolution of the electron energy during the interaction to determine the resulting polarization spectrum for the emitted photons \cite{Ridgers_17}:

\begin{equation}
    \frac{d\gamma}{dt} = -\frac{2\alpha_f c}{3\lambdabar_c}\chi_e^2(t)G(\chi_e)\;,
\label{radiation_reaction_equation}
\end{equation}

Here, $G(\chi_e)$ is the Gaunt factor accounting for quantum radiation corrections, which can be approximated as $G(\chi_e)\approx[1 + 4.8(1 + \chi_e)\rm{ln}(1 + 1.7\chi_e) + 2.44\chi_e^2]^{-2/3}$. The electrons emit polarized x-rays with a spectrum $S(\varepsilon_\gamma, \pmb{\xi})$, where $\varepsilon_\gamma$ is the photon energy and $\pmb{\xi}$ is the polarization state. Since the participating electrons in the NLCS process are initially unpolarized, the emitted x-ray photons can only be linearly polarized. Thus, we characterize the polarization state using the degree of linear polarization, $\xi_3$, where $\xi_3 \in \{-1, +1\}$. Here, $\xi_3 = +1$ corresponds to photons fully polarized along the laser polarization direction $\hat{\pmb{o}}_1$, while $\xi_3 = -1$ corresponds to full polarization along $\hat{\pmb{o}}_2$. The total radiation spectrum is then obtained by integrating the polarization-resolved NLCS spectrum at each time step over the full interaction duration, $\tau_0$:

\begin{equation}
S(\varepsilon_\gamma, \xi_3) = C_0\int_0^{\tau_{0} } dt F_{\text{NLC}}(\chi_e(t),\varepsilon_\gamma,\xi_3).
\end{equation}

The explicit form of the polarization-resolved NLCS spectrum, $F_{\text{NLC}}(\chi_e,\, \varepsilon_\gamma,\, \xi_3)$, is provided in Eqs.~\ref{eq:NLC_stokes_F0} and \ref{eq:NLC_stokes_F3}. The total number of photons emitted with energy above a threshold $\varepsilon_{th}$, denoted $N_\gamma(\varepsilon_{th})$, can be calculated by integrating the photon spectrum above $\varepsilon_{th}$ and multiplying by the estimated number of interacting electrons $N_e$:

\begin{equation}
N_\gamma(\varepsilon_{th}, \xi_3) = N_e \times \int_{\varepsilon_{th}}^{\infty} d\varepsilon_\gamma\ S(\varepsilon_\gamma, \xi_3). 
\end{equation}

The expected degree of linear polarisation can be calculated as:

\begin{equation}
\langle\xi_3\rangle(\varepsilon_\gamma) = \frac{S(\varepsilon_\gamma, +1)-S(\varepsilon_\gamma, -1)}{S(\varepsilon_\gamma, +1)+S(\varepsilon_\gamma, -1)}.
\end{equation}

Also, the expected degree of linear polarisation above the energy threshold $\varepsilon_{th}$ is:

\begin{equation}
\langle\xi_3\rangle(\varepsilon_{th}) = \frac{N(\varepsilon_{th}, +1)-N(\varepsilon_{th}, -1)}{N(\varepsilon_{th}, +1)+N(\varepsilon_{th}, -1)}.
\end{equation}

%Plot for the number of generated photons above energy $E$. We assume the number of electrons during the interaction, Laser spot size. We use an aluminum target, whose density is about $450n_c$. Quantum parameter $\chi$ is estimated using equations (from Riders JPP 2017): 

\section{Comparison of hard photon polarisation model to PIC simulations}

To evaluate our model’s predictions for the degree of polarization of emitted NLCS x-ray photons, we carried out particle-in-cell (PIC) simulations using the {\small{OSIRIS}} code \cite{Fonseca_Osiris_note, Fonseca_Osiris_PPCF}. The {\small{OSIRIS}} framework has recently been extended to account for the polarization of high-energy photons. Additionally, our simulations incorporate a spin- and polarization-resolved pair production process. However, under the parameters considered in this study, pair production is negligible. The implementation of these features will be briefly outlined below; further details can be found in Ref.~\cite{Qian_23}.

\subsection{PIC with polarised hard photon emission}

{\small{OSIRIS}} updates each electron’s position, momentum, and spin within the PIC loop by employing the classical Lorentz pusher for particle dynamics and a 'spin pusher' that integrates the Thomas–Bargmann–Michel–Telegdi (T-BMT) equation for spin evolution. Afterward, the code computes each electron’s quantum parameter $\chi_e$ and the direction of the magnetic field in the particle rest frame, $\pmb{\hat\beta}$. The electron’s spin component along $\pmb{\hat\beta}$, denoted $\pmb{S}_\beta$, together with $\chi_e$, is used to evaluate the spin-dependent NLCS emission rate.

A random number $r_1 \in [0,1]$ is drawn and compared to the emission probability to stochastically determine whether a radiation event occurs. If an emission takes place, an additional random number $r_2 \in [0,1]$ is used, along with a precomputed inverse cumulative distribution table for the photon spectrum, to sample the photon energy $\varepsilon_\gamma$. Under the ultra-relativistic approximation, the emitted photon’s momentum direction is assumed to coincide with that of the parent electron.

To determine the lepton’s spin state following emission, the code calculates the expected spin vector $\pmb S_{SQA}$, which depends on the initial spin state, $\chi_e$, the spin basis vectors $(\pmb{\hat \alpha}, \pmb{\hat \beta}, \pmb{\hat \kappa})$, and $\varepsilon_\gamma$. The explicit expression for the expected spin direction is given in Eq.~\ref{eq:SQA_spin}. Since $|\pmb S_{SQA}|$ is generally less than unity, spin quantization is enforced as follows: a third random number $r_3 \in [0,1]$ is drawn; if $r_3 < (1 + |\pmb S_{SQA}|)/2$, the spin aligns with $\pmb S_{SQA}/|\pmb S_{SQA}|$, otherwise it is anti-aligned.

Subsequently, the initial and final electron spin states, the spin basis vectors $(\pmb{\hat \alpha}, \pmb{\hat \beta}, \pmb{\hat \kappa})$, $\chi_e$, and $\varepsilon_\gamma$ are used to compute the expected Stokes vector $\pmb \xi$ describing the polarization state of the emitted photon. The Stokes vector is similarly projected to a pure state: a random number $r_4$ is drawn, and the polarization vector is set to $\pm \pmb \xi/|\pmb \xi|$ via a Monte Carlo selection, with $r_4 < (1 + |\pmb \xi|)/2$ yielding alignment and the complement yielding anti-alignment. Finally, consistent with the conditional probability structure of our Monte Carlo algorithm \cite{Chen_1995_CAIN, Chen_PRD_2022}, the electron’s spin is also updated carefully for cases when an emission event does not occur.

\subsection{PIC simulation results for hard photon polarisation}

\begin{figure}[h!]
\centering
\includegraphics[scale=0.11]{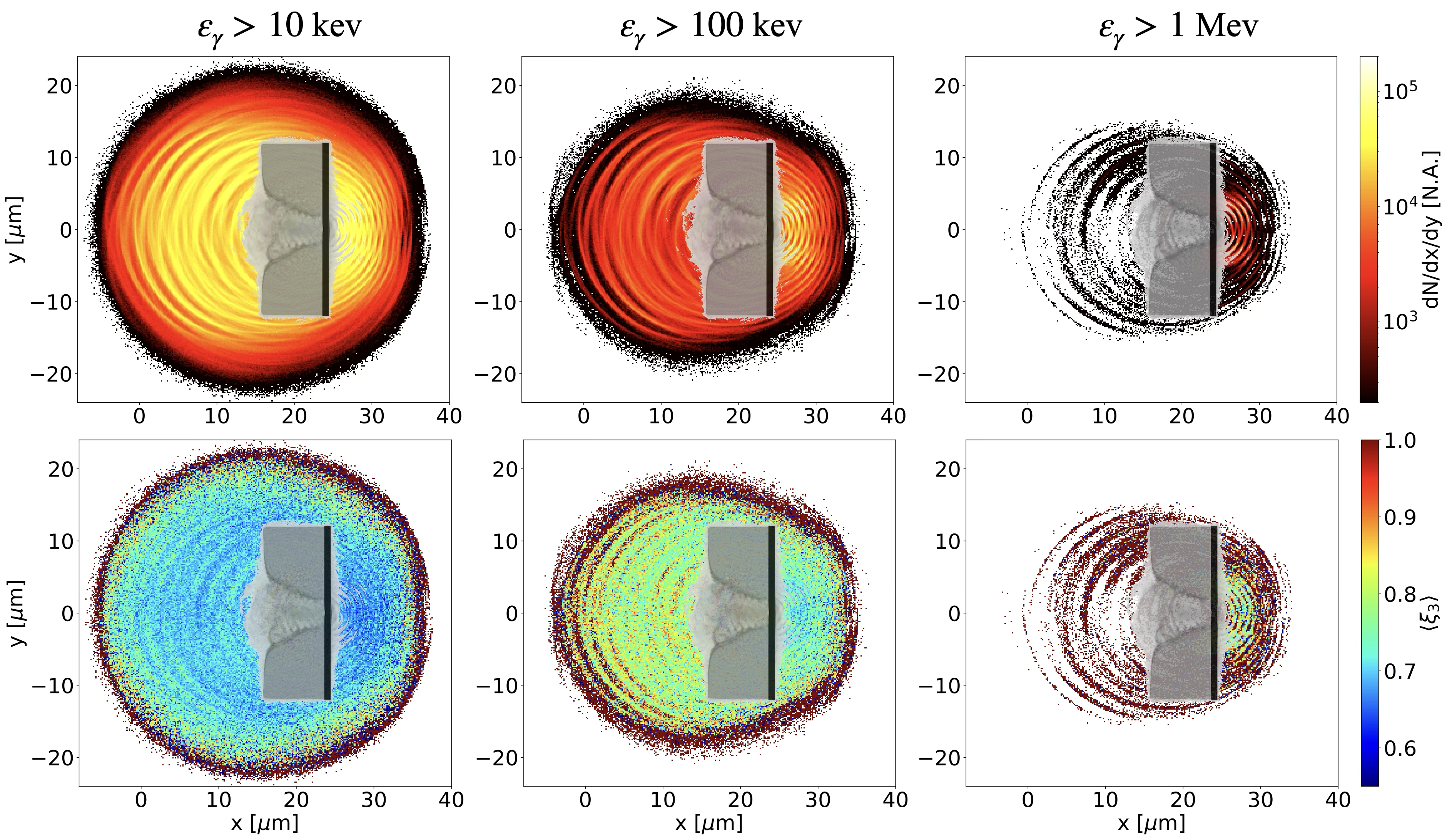}
\caption{Top row: Photon density in the $x$-$y$ simulation plane for photons with energies above (a) 10 keV, (b) 100 keV, and (c) 1 MeV, shown after the laser pulse is fully reflected from the target. Bottom row: Spatial distribution of the photon linear polarization degree, $\langle\xi_3\rangle$, for the same energy thresholds—(d) 10 keV, (e) 100 keV, and (f) 1 MeV. Here, $\langle\xi_3\rangle = 1$ corresponds to photons fully polarized along the laser's polarization direction. The grayscale overlay indicates the target electron density distribution.}
\label{fig:photon_dens_pol}
\end{figure}

\begin{figure}[h!]
\centering
\includegraphics[scale=0.12]{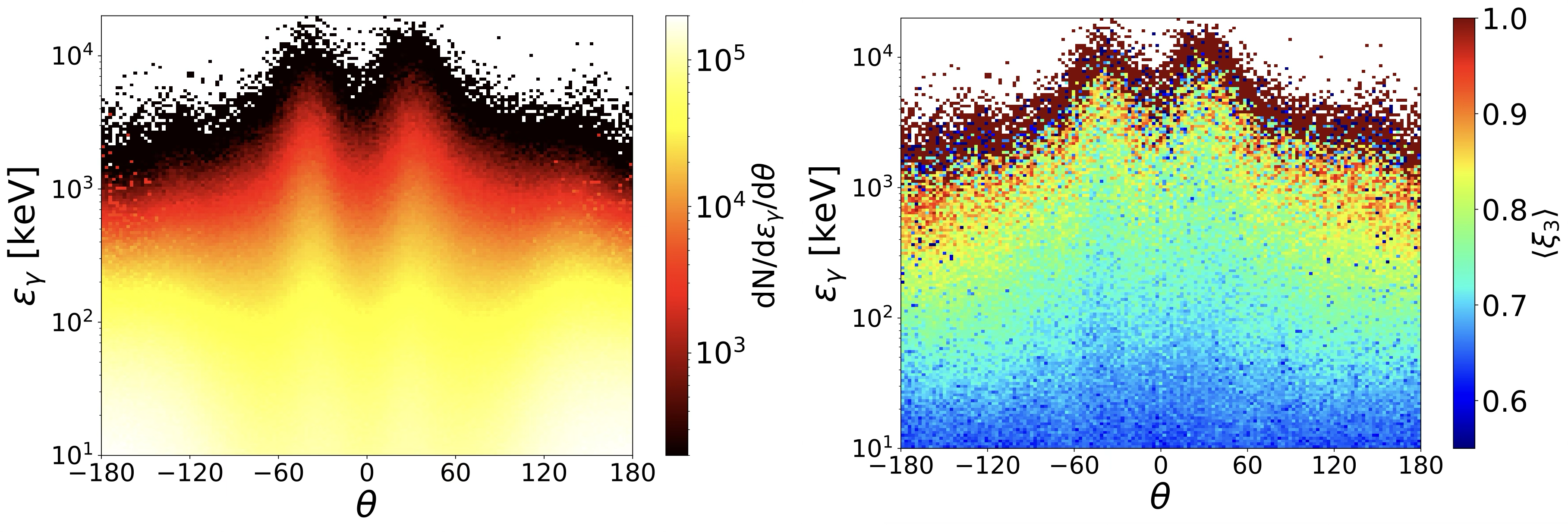}
\caption{(a) Angle-resolved photon density distribution, $\frac{d^2N}{d\theta,dE}$; (b) corresponding distribution of the photon linear polarization degree, $\langle\xi_3\rangle$. Both angular maps correspond to the same simulation as shown in Fig.~\ref{fig:photon_dens_pol}.}
\label{fig:photon_angle_ene}
\end{figure}

To benchmark our model from Sec.~\ref{simple_model}, we performed 2D spin- and polarization-resolved PIC simulations using a laser pulse incident on a $0.8$,$\mu$m-thick aluminum foil with electron density $n_p = 450n_c$. An $8\,\mu$m-long pre-plasma region with $n_p = 5n_c$ is present in front of the target. The incident laser has normalized vector potentials $a_0 = 15,\ 20,\ 25,\ 30,\ 40$, wavelength $\lambda = 0.8\,\mu$m, focal spot size $w_0 = 4\,\mu$m, and duration $\tau_0 = 20\,fs$. The laser is linearly polarized along the $y$-axis and propagates in the $+x$-direction in the simulation. 

We first present {\small{OSIRIS}} simulation results for $a_0 = 25$. Figure \ref{fig:photon_dens_pol} shows both the 2D photon density and the spatial map of the average degree of linear polarization ($\langle\xi_3\rangle$) for x-ray photons with energies above 10 keV, 100 keV, and 1 MeV, respectively. For reference, the electron density distribution is overlaid on the photon maps. Figure \ref{fig:photon_angle_ene} presents the angular-energy resolved photon density and degree of linear polarization. These results reveal that lower-energy x-ray photons (10–100 keV) are more uniformly distributed in the $x$-$y$ plane and tend to propagate in the $-x$-direction, while higher-energy x-rays ($>100$ keV) are more likely to travel in the $+x$-direction. Additionally, the degree of linear polarization increases with photon energy.

\begin{figure}[h!]
\centering
\includegraphics[scale=0.45]
{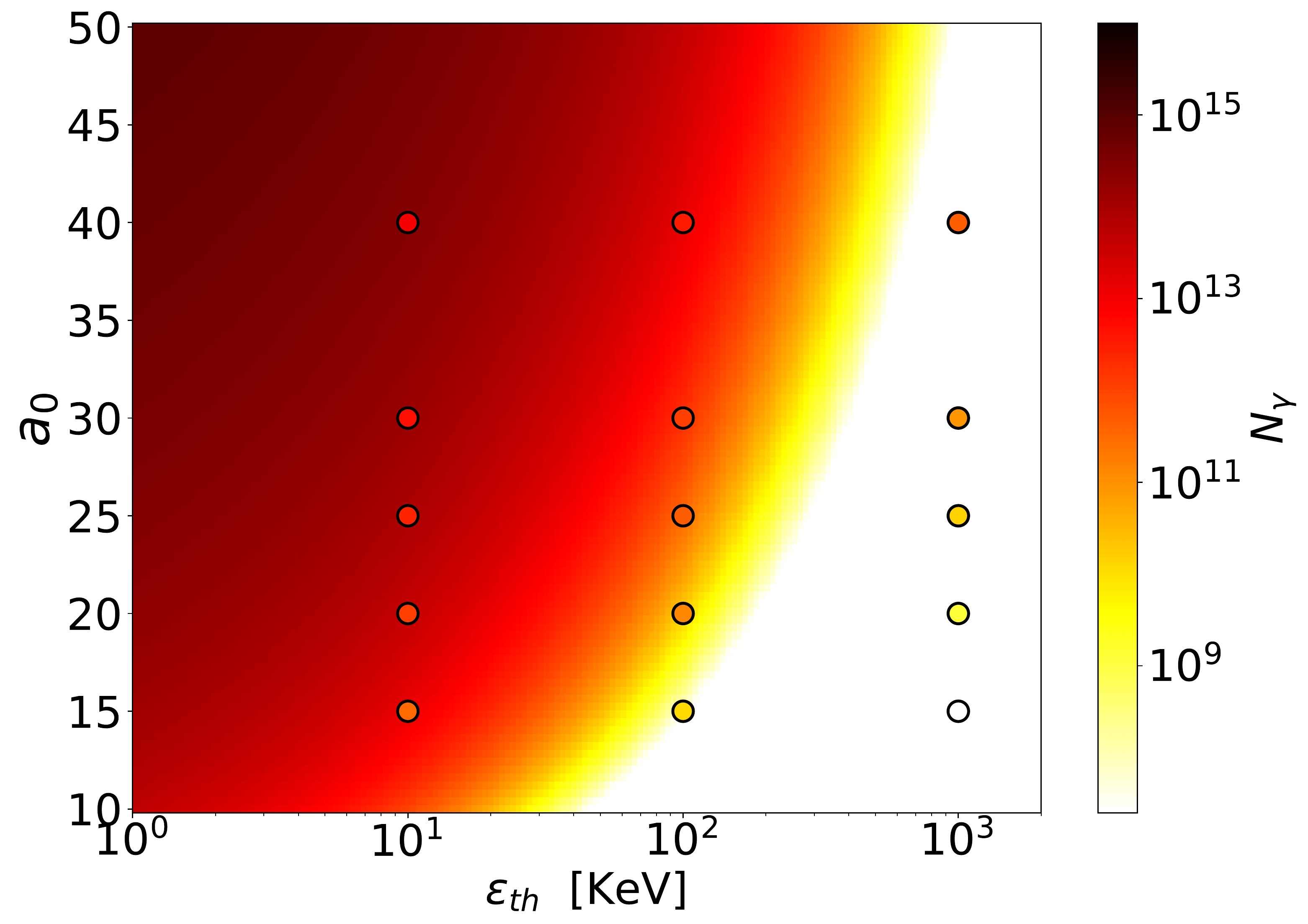}
\caption{Number of NLCS x-ray photons with energies above $\varepsilon_{th}$ produced in laser–solid target interactions at various laser intensities ($a_0$).}
\label{fig:photon_number_ana_sim}
\end{figure}

\begin{figure}[h!]
\centering
\includegraphics[scale=0.5]
{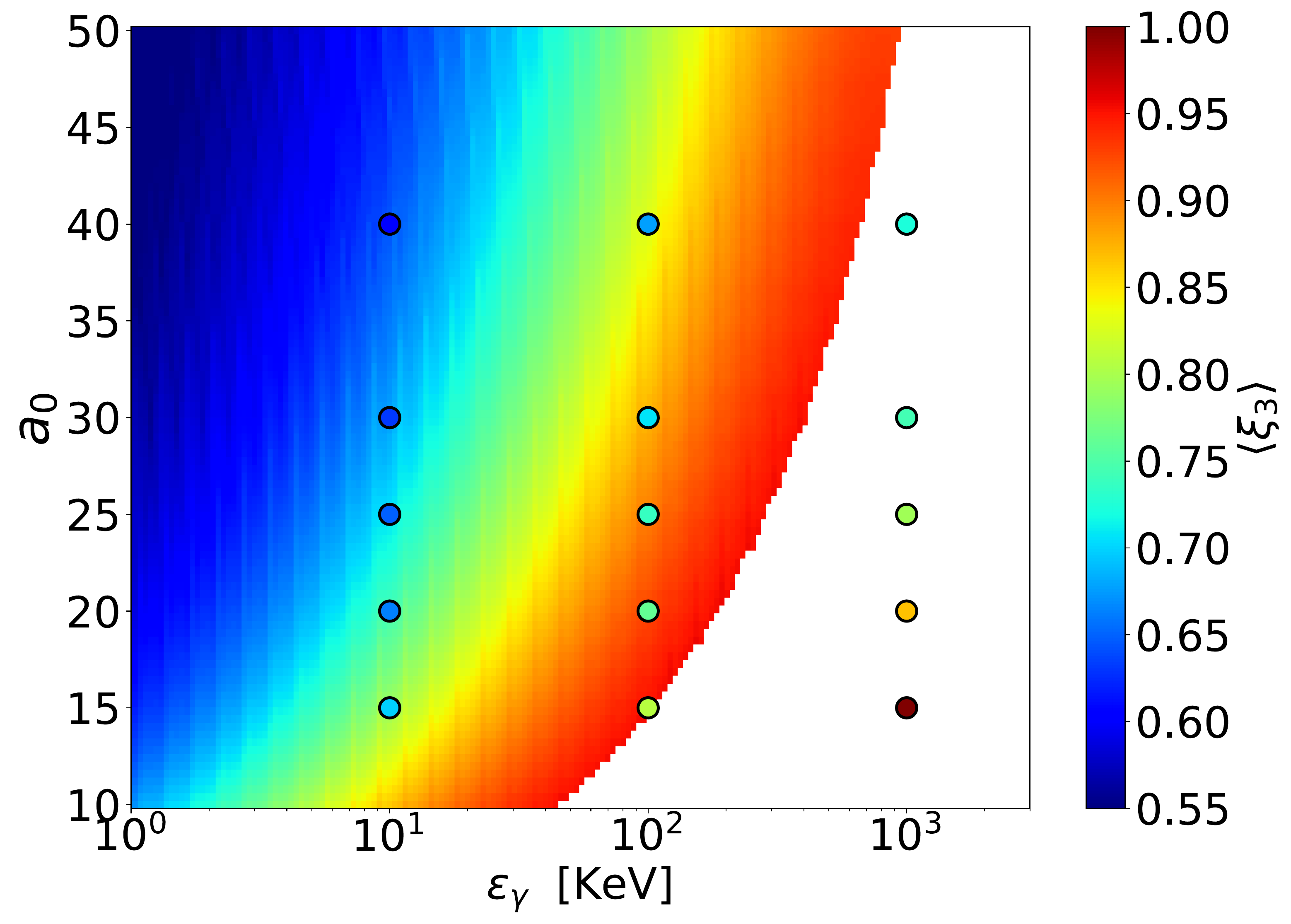}
\caption{Spectrum of NLCS x-ray photon polarization in laser–solid target interactions, shown as a function of $a_0$ and emitted photon energy.}
\label{fig:photon_polarisaiton_ana_sim}
\end{figure}

\begin{figure}[h!]
\centering
\includegraphics[scale=0.5]
{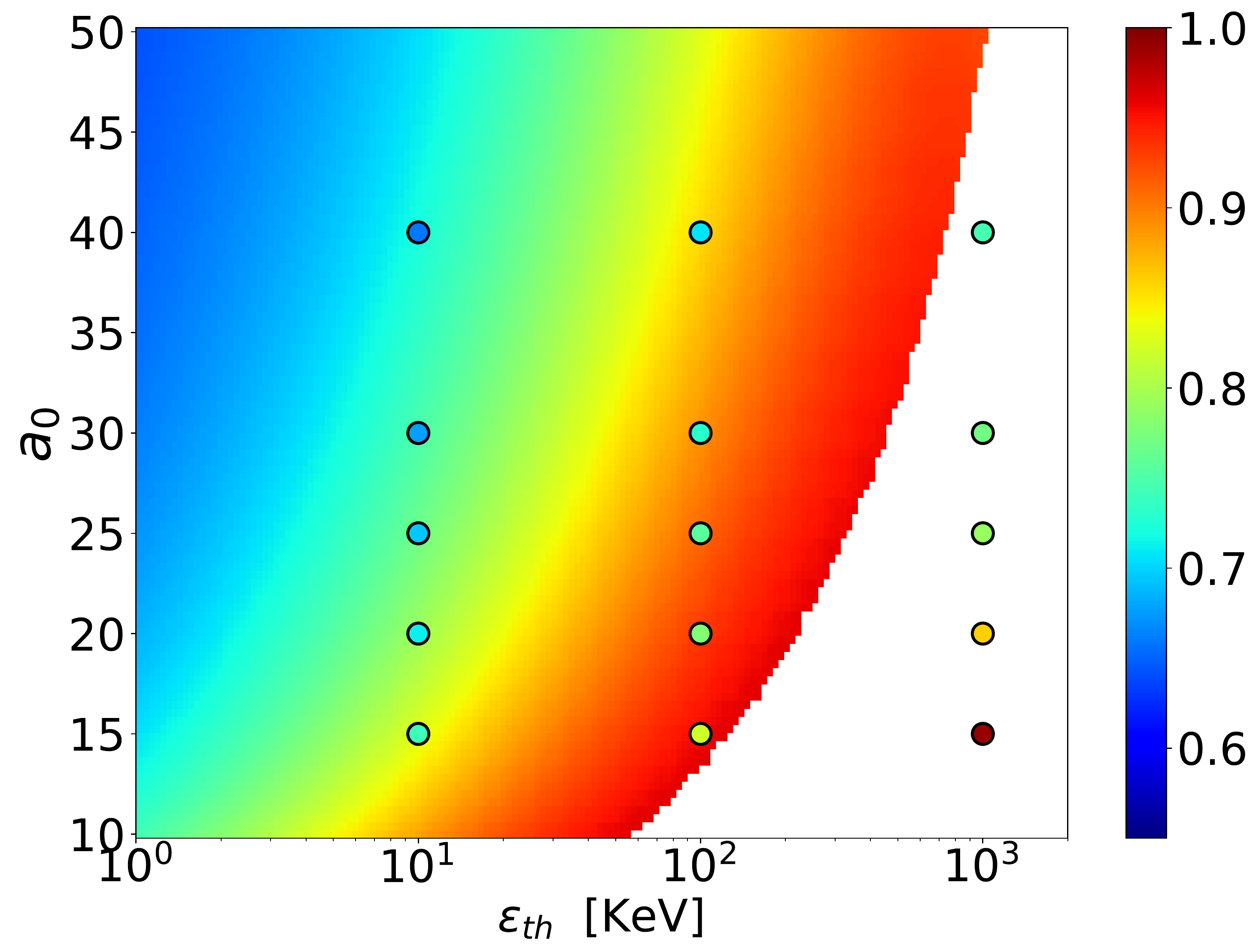}
\caption{Degree of polarization of NLCS x-ray photons with energies above $\varepsilon_{th}$ in laser–solid target interactions at varying laser intensities ($a_0$).}
\label{fig:photon_polarisaiton_int_ana_sim}
\end{figure}

Figures \ref{fig:photon_number_ana_sim} and \ref{fig:photon_polarisaiton_int_ana_sim} show the predicted photon yield and polarization degree for photons above various threshold energies $\varepsilon_{th}$, using the simplified model from Sec.\ref{simple_model}. Figure \ref{fig:photon_polarisaiton_ana_sim} presents the predicted polarization spectrum. We compare the simulated photon yield and polarization degree for varying $a_0$ in Figs.~\ref{fig:photon_number_ana_sim}, \ref{fig:photon_polarisaiton_ana_sim}, and \ref{fig:photon_polarisaiton_int_ana_sim}; the results are also tabulated in Tables~\ref{table:sim_yield_table}, \ref{table:sim_pol_table}, and \ref{table:sim_pol_int_table}.

Although the simulations and the simplified model are based on similar physical principles, their setups differ considerably, resulting in quantitative differences in both photon yield and polarization degree. Nevertheless, both approaches display consistent qualitative trends within the explored intensity range ($a_0 = 10$–$50$): the linear polarization of emitted x-rays decreases with increasing laser $a_0$ and increases with photon energy. Notably, for photons with energies above 10,keV, the polarization degree remains above $65\%$ and can exceed $80\%$ at higher energies. These results provide valuable insights for future experimental design.

\begin{table*}[th!]
  \centering

\begin{tabular}{c c c c c c}
 $\ \ \text{Energy}\ \varepsilon_{th}\ $  & $\ \ a_0= 15\ \ $ &$\ \ a_0= 20\ \ $ &$\ \ a_0= 25\ \ $ &$\ \ a_0= 30\ \ $ & $\ \ a_0= 40\ \ $ \\
\hline
$>10$\thinspace keV & $N_{\gamma} = 3.19\times 10^{11}$ & $1.1\times 10^{12}$ & $2.5\times 10^{12}$ & $4.6\times 10^{12}$  & $9.7\times 10^{12}$\\ 

$>100$\thinspace keV & $N_{\gamma} = 1.2\times 10^{10}$ & $1.25\times 10^{11}$ & $4.4\times 10^{11}$& $1.16\times 10^{11}$  & $3.35\times 10^{11}$\\ 

$>1$\thinspace MeV & $N_{\gamma} = 7\times 10^{6}$ & $1.2\times 10^{9}$ & $1.5\times 10^{10}$& $8.2\times 10^{10}$  & $4.36\times 10^{11}$\\ 

\end{tabular}

\caption{Simulation results shown in Fig.~\ref{fig:photon_number_ana_sim} for NLCS x-ray photon number above energy $\varepsilon_{th}$ in a laser solid target interaction with different laser intensity $a_0$.}
\label{table:sim_yield_table}
\end{table*}

\begin{table*}[th!]
  \centering

\begin{tabular}{c c c c c c}
 $\ \ \text{Energy}\ \varepsilon_{\gamma}\ $  & $\ \ a_0= 15\ \ $ &$\ \ a_0= 20\ \ $ &$\ \ a_0= 25\ \ $ &$\ \ a_0= 30\ \ $ & $\ \ a_0= 40\ \ $ \\
\hline
$=10 \text{ keV}$ & $\langle\xi_3\rangle = 70\%$ & $67.6\%$ & $64.9\%$ & $63.1\%$  & $60.5\%$\\ 

$=100 \text{ keV}$ & $\langle\xi_3\rangle = 81\%$ & $76.8\%$ & $73.7\%$& $70.8\%$  & $67.8\%$\\ 

$=1  \text{ MeV}$ & $\langle\xi_3\rangle = 100\%$ & $87.1\%$ & $79.6\%$& $74.5\%$  & $72.3\%$\\ 

\end{tabular}

\caption{Simulation results in Fig. 4 for the spectrum of NLCS x-ray photon polarization in laser–solid target interactions, as a function of $a_0$ and emitted photon energy.}
\label{table:sim_pol_table}
\end{table*}

\begin{table*}[th!]
  \centering

\begin{tabular}{c c c c c c}
 $\ \ \text{Energy}\ \varepsilon_{th}\ $  & $\ \ a_0= 15\ \ $ &$\ \ a_0= 20\ \ $ &$\ \ a_0= 25\ \ $ &$\ \ a_0= 30\ \ $ & $\ \ a_0= 40\ \ $ \\
\hline
$>10\text{ keV}$ & $\langle\xi_3\rangle = 74\%$ & $73.1\%$ & $69.5\%$ & $67.8\%$  & $66\%$\\ 

,$>100\text{ keV}$ & $\langle\xi_3\rangle = 81.9\%$ & $79.7\%$ & $75.6\%$& $73\%$  & $71\%$\\ 

$>1\text{ MeV}$ & $\langle\xi_3\rangle = 99\%$ & $89.5\%$ & $79\%$& $77\%$  & $74.6\%$\\ 

\end{tabular}

\caption{Simulation results shown in Fig.~\ref{fig:photon_polarisaiton_int_ana_sim} for the degree of polarization of NLCS x-ray photons with energies above  $\varepsilon_{th}$ in a laser solid target interaction with different laser intensity $a_0$.}
\label{table:sim_pol_int_table}
\end{table*}
                         
\section{Conclusions}
By employing a newly developed emission module in the {\small{OSIRIS}} PIC code, we demonstrate that keV photons generated via nonlinear Compton scattering (NLCS) in $>1$ PW laser-plasma interactions are highly polarized, with degrees of polarization exceeding $80\%$. This stands in clear contrast to x-rays produced by other mechanisms, such as bremsstrahlung, which typically exhibit much lower polarization. As a result, x-ray polarimetry offers a promising diagnostic for distinguishing NLCS x-ray flashes from background sources.

\section{Acknowledgment}
This work was supported by the National Science Foundation grant 2108075, NSF-GACR collaborative grant 2206059 from the NSF, and Czech Science Foundation Grant No. 22-42963L. C. P. R. was supported by UK EPSRC grant number EP/V049461/1 and
funding from ELI-ERIC. T.G. and M.V. are supported by FCT (Portugal) Grants No. CEECIND/04050/2021 and No. PTDC/FISPLA/3800/2021. The authors would like to acknowledge the OSIRIS Consortium, consisting of UCLA and IST (Lisbon, Portugal), for providing access to the OSIRIS 4.0 framework. Work supported by NSF ACI-1339893.

\section{Data Availability}

The data required to reproduce the results presented in this paper is available at

\ \\

\bibliography{gamma_polarisation}

\end{document}